\documentclass[a4paper,10pt,twoside]{cpc-hepnp}

\usepackage{multicol}
\usepackage{graphicx}
\usepackage{booktabs}
\usepackage{amssymb,bm,mathrsfs,bbm,amscd}
\usepackage[tbtags]{amsmath}
\usepackage{lastpage}
\def\ka{\left.\mid 11 \right\rangle_c}
\def\kb{\left.\mid 88 \right\rangle_c}
\def\kap{\left.\mid 1'1' \right\rangle_c}

\begin{document}

\fancyhead[co]{\footnotesize A. Valcarce et al: Molecular and compact four-quark states}

\footnotetext[0]{Received 21 October 2009}

\title{Molecular and compact four-quark states\thanks{This work has been 
partially funded by Ministerio de Ciencia y Tecnolog\'{\i}a
under Contract No. FPA2007-65748, by EU FEDER, and by Junta de Castilla y Le\'{o}n
under Contracts No. SA016A17 and GR12}}

\author{%
      A. Valcarce$^{1;1)}$\email{valcarce@usal.es}%
\quad J. Vijande$^{1;2)}$\email{javier.vijande@uv.es}%
}
\maketitle

\address{%
1~Departamento de F\'{\i}sica Fundamental, Universidad de Salamanca,Salamanca, Spain\\
2~Departamento de F\'{\i}sica At\'{o}mica, Molecular y Nuclear, Universidad de Valencia (UV)
and IFIC (UV-CSIC), Valencia, Spain\\
}

\begin{abstract}
We study charmonium ($c\bar c n\bar n$), bottomonium ($b\bar b n\bar n$)
and exotic ($cc\bar n\bar n$ and $bb\bar n\bar n$)
four-quark states by means of a standard non-relativistic quark potential model.
We look for possible bound states. Among them we are able to distinguish between
meson-meson molecules and compact four-quark states.
\end{abstract}



\begin{multicols}{2}

\section{Introduction}
The discoveries on several fronts of unusual
charmonium states like $X(3872)$ and $Y(4260)$ and
open-charm mesons with unexpected masses like
$D_{sJ}^*(2317)$ and $D^*_0(2308)$,
have re-invigorated the study of the hadron spectra.
Their anomalous nature has triggered
several interpretations, among them, the existence of
compact four-quark states or meson-meson molecules. This challenging situation
resembles the long-standing problem of the light-scalar
mesons, where it has been suggested that some resonances
may not be ordinary $q \bar q$ states, though
there is little agreement on what they actually are~\cite{Ams04}.
In this case, four-quark states have been justified to coexist
with $q\bar q$ states because they can couple to $J^{PC}=0^{++}$ without
orbital excitation~\cite{Jaf05}.

Four-quark systems present a richer color structure than standard baryons or mesons.
Although the color wave function for ordinary mesons and baryons leads to a single vector,
working with four--quark states there are different vectors driving to a singlet color state
out of colorless or colored quark-antiquark two-body components. Thus, in dealing with four--quark states
an important question is whether we are in
front of a colorless meson--meson molecule or a compact
state (i.e., a system  with two-body colored components).
Whereas the first structure would be natural in the naive quark model,
the second one would open a new area in hadron spectroscopy.

We have derived the necessary formalism to evaluate the
physical channels probability(singlet--singlet color states) in an
arbitrary four--quark
wave function. For this purpose we have expanded any hidden--color vector of the
four--quark state color basis, i.e., vectors with non--singlet internal color couplings,
in terms of singlet--singlet color vectors. 
Such a procedure gives rise to a 
wave function expanded in terms of color singlet-singlet
nonorthogonal vectors, where the determination of the probability of physical
channels becomes cumbersome. However, it allows to differentiate among
unbound, compact and molecular four-quark states. 

\section{Formalism}
There are three different ways of coupling two quarks and two antiquarks
into a colorless state:
\end{multicols}
\begin{subequations}
\begin{eqnarray}
\label{eq1a}
[(q_1q_2)(\bar q_3\bar q_4)]&\equiv&\{|\bar 3_{12}3_{34}\rangle,|6_{12}\bar 6_{34}\rangle\}\equiv\{|\bar 33\rangle_c^{12},
|6\bar 6\rangle_c^{12}\}\\
\label{eq1b}
[(q_1\bar q_3)(q_2\bar q_4)]&\equiv&\{|1_{13}1_{24}\rangle,|8_{13} 8_{24}\rangle\}\equiv\{|11\rangle_c,|88\rangle_c\}\\
\label{eq1c}
[(q_1\bar q_4)(q_2\bar q_3)]&\equiv&\{|1_{14}1_{23}\rangle,|8_{14} 8_{23}\rangle\}\equiv\{|1'1'\rangle_c,|8'8'\rangle_c\}\,,
\end{eqnarray}
\label{eq1}
\end{subequations}
\begin{multicols}{2}
each forms an orthonormal basis. Each coupling scheme allows us to
define a color basis where the four--quark problem can be solved.
The first basis, Eq.~(\ref{eq1a}),
being the most suitable one to deal with the Pauli principle is made
entirely of vectors containing hidden--color components. The other two, Eqs.~(\ref{eq1b}) and~(\ref{eq1c}),
are hybrid bases containing singlet--singlet (physical) and octet--octet (hidden--color)
vectors.

An arbitrary four-quark wave function can be expressed in terms of physical 
components, singlet-singlet color states, by means of an infinite expansion
than can be resumed as~\cite{Vij09},
\begin{eqnarray}
|\Psi\rangle  & = & \frac{1}{2} \left( P\hat Q + \hat Q P \right) \frac{1}{1-\cos^2\alpha}  |\Psi\rangle \nonumber \\
        & + & \frac{1}{2} \left( \hat P Q + Q \hat P \right) \frac{1}{1-\cos^2\alpha} |\Psi\rangle\,.
\end{eqnarray}
Thus, one obtains two hermitian operators that are well--defined
projectors on the two physical singlet--singlet color states
\begin{eqnarray}
{\cal P}_{\ka} & =&  \left( P\hat Q + \hat Q P \right) \frac{1}{2(1-\cos^2\alpha)}
\nonumber \\
{\cal P}_{\kap} & =&  \left( \hat P Q + Q \hat P \right) \frac{1}{2(1-\cos^2\alpha)} \,.
\label{tt}
\end{eqnarray}
Thus, given an arbitrary state $|\Psi\rangle$ its projection on a particular subspace $E$ is given
by $|\Psi\rangle|_E=P_E|\Psi\rangle$. Then, the probability of finding such an state on this subspace is
\begin{equation}
_E|\langle\Psi|\Psi\rangle|_E=\langle\Psi|P_E^{\dagger}P_E|\Psi\rangle=\langle\Psi|P_E^2|\Psi\rangle=\langle\Psi|P_E|\Psi\rangle\,.
\end{equation}
Therefore, once the projection operators have been constructed [Eq.~(\ref{tt})], the probabilities for finding
singlet--singlet components are given by,
\begin{eqnarray}
P^{\mid\Psi\rangle}({[11]})&=&\left\langle\Psi\mid{\cal P}_{\ka}\mid\Psi\right\rangle\nonumber\\
P^{\mid\Psi\rangle}({[1'1']})&=&\left\langle\Psi\mid{\cal P}_{\kap}\mid\Psi\right\rangle\, .
\end{eqnarray}
Using Eq.~(\ref{tt}) it can be easily checked that $P^{\mid\Psi\rangle}({[11]})+
P^{\mid\Psi\rangle}({[1'1']})=1$, where
\end{multicols}
\begin{eqnarray}
P^{\mid\Psi\rangle}({[11]})&=&\frac{1}{2(1-\cos^2\alpha)}
\left[ \left\langle\Psi\mid P\hat Q \mid\Psi\right\rangle +
\left\langle\Psi\mid \hat Q P \mid\Psi\right\rangle\right] \nonumber \\
P^{\mid\Psi\rangle}({[1'1']})&=&\frac{1}{2(1-\cos^2\alpha)}
\left[ \left\langle\Psi\mid \hat P Q \mid\Psi\right\rangle +
\left\langle\Psi\mid Q \hat P \mid\Psi\right\rangle\right] \, .
\end{eqnarray}
\begin{multicols}{2}

\section{Results}
The stability of a four--quark state can be analyzed in terms of $\Delta_E$,
the energy difference between its mass and that of the
lowest two-meson threshold,
\begin{equation}
\label{delta}
\Delta_E=E_{4q}-E(M_1,M_2)\, ,
\end{equation}
where $E_{4q}$ stands for the four--quark energy and $E(M_1,M_2)$ for the energy of the
two-meson threshold. Thus, $\Delta_E<0$ indicates all fall-apart decays
are forbidden, and therefore one has a proper bound state. $\Delta_E\ge 0$
will indicate that the four--quark solution corresponds to
an unbound threshold (two free mesons). Thus, an energy above the threshold
would simply mean that the system is unbound within our variational
approximation, suggesting that the minimum of
the Hamiltonian is at the two-meson threshold.

Another helpful tool analyzing the structure of a four--quark state is
the value of the root mean square radii:
$\langle x^2\rangle^{1/2}$, $\langle y^2\rangle^{1/2}$, and $\langle z^2\rangle^{1/2}$.
Bound four--quark states can be distinguished from
two free mesons by means of their root mean square radius
\begin{equation}
{\rm RMS}_{4q(2q)}= \left(\frac{\sum_{i=1}^{4(2)} m_i \langle (r_i-R)^2\rangle}{\sum_{i=1}^{4(2)} m_i}\right)^{1/2}
\,,
\end{equation}
and in particular, their corresponding ratio,
\begin{equation}
\label{delta-r}
\Delta_R=\frac{{\rm RMS}_{4q}}{{\rm RMS}_{M_1}+{\rm RMS}_{M_2}}\,,
\end{equation}
where ${\rm RMS}_{M_1}+{\rm RMS}_{M_2}$ stands for the sum of the radii of the mesons
corresponding to the lowest threshold.

We show in Table~\ref{t1} the results obtained for
several different four--quark states in the bottom
and charm sectors. One can see how
independently of their binding energy, all of
them present a sizable octet-octet
component when the wave function is expressed in the (\ref{eq1b}) coupling.
Let us first of all concentrate on the
two unbound states, $\Delta_E > 0$, one with $S_T=0$ and one with $S_T=1$, given
in Table~\ref{t1}. The octet-octet component of basis (1b) can be expanded in terms of
the vectors of basis (1c) as explained in the previous section. Thus,
once expressions (6) are considered one finds that
the probabilities are concentrated into a single physical channel, $MM$ or $MM^*$.
In other words, the octet-octet component of the basis (1b) or (1c) is a
consequence of having identical quarks and antiquarks. Thus, four-quark
unbound states are represented by
two isolated mesons. This conclusion is strengthened when studying
the root mean square radii,
leading to a picture where the two quarks and the two antiquarks are far
away, $\langle x^2\rangle^{1/2}\gg 1$ fm and $\langle y^2\rangle^{1/2}\gg 1$ fm,
whereas the quark-antiquark pairs are located at a typical distance
for a meson, $\langle z^2\rangle^{1/2}\le 1$ fm.

\end{multicols}
\begin{center}
\tabcaption{\label{t1}Four--quark state properties for selected quantum numbers.
All states have positive parity and total orbital angular momentum $L=0$.
Energies are given in MeV and distances in fm. The notation
$M_1M_2\mid_{\ell}$ stands for mesons $M_1$ and $M_2$ with a relative orbital
angular momentum $\ell$. $P[| \bar 3 3\rangle_c^{12}(| 6\bar 6\rangle_c^{12})]$ stands for the
probability of the $3\bar 3(\bar 6 6)$ components given in Eq.~(\ref{eq1a}) and $P[\ka(\kb)]$ for the
$11(88)$ components given in Eq.~(\ref{eq1b}). $P_{MM}$, $P_{MM^*}$, and $P_{M^*M^*}$ stand for the
probability of two pseudoscalar, pseudoscalar-vector or two vector mesons.}
\vspace{-3mm}
\footnotesize
\begin{tabular*}{170mm}{@{\extracolsep{\fill}}|c|ccccc|}
\toprule $(S_T,I)$                       & (0,1)          &  (1,1)          & (1,0)         & (1,0)          & (0,0) \\
Flavor                          &$cc\bar n\bar n$&$cc\bar n\bar n$&$cc\bar n\bar n$&$bb\bar n\bar n$&$bb\bar n\bar n$\\
\hline
Energy                          & 3877           &  3952           & 3861          & 10395          & 10948 \\
Threshold                       & $DD\mid_S$     &  $DD^*\mid_S$   & $DD^*\mid_S$   & $BB^*\mid_S$   &  $B_1B\mid_P$\\
$\Delta_E$                      & +5             &  +15            & $-76$         & $-$217         &  $-153$ \\
\hline
$P[| \bar 3 3\rangle_c^{12}]$   & 0.333          &  0.333          & 0.881         & 0.974          &  0.981 \\
$P[| 6 \bar 6\rangle_c^{12}]$   & 0.667          &  0.667          & 0.119         & 0.026          &  0.019 \\
\hline
$P[\ka]$                        & 0.556          &  0.556          & 0.374         & 0.342          &  0.340 \\
$P[\kb]$                        & 0.444          &  0.444          & 0.626         & 0.658          &  0.660 \\
\hline
$P_{MM}$                        & 1.000          &  $-$            & $-$           & $-$            &  0.254 \\
$P_{MM^*}$                      & $-$            &  1.000          & 0.505         & 0.531          &  $-$ \\
$P_{M^*M^*}$                    & 0.000          &  0.000          & 0.495         & 0.469          &  0.746 \\
\hline
$\langle x^2\rangle^{1/2}$      & 60.988         &  13.804         & 0.787         & 0.684          &  0.740 \\
$\langle y^2\rangle^{1/2}$      & 60.988         &  13.687         & 0.590         & 0.336          &  0.542 \\
$\langle z^2\rangle^{1/2}$      & 0.433          &  0.617          & 0.515         & 0.503          &  0.763 \\
$RMS_{4q}$                      & 30.492         &  6.856          & 0.363         & 0.217          &  0.330 \\
$\Delta_R$                      & 69.300         & 11.640          &0.799          & 0.700          &  0.885 \\
\bottomrule
\end{tabular*}%
\end{center}

\begin{multicols}{2}

Let us now turn to the bound states shown
in Table \ref{t1}, $\Delta_E < 0$, one in the charm sector and two in the bottom one.
In contrast to the results obtained for unbound states, when the octet-octet
component of basis (1b) is expanded in terms of the vectors of basis (1c),
equations (6) indicate a picture where the probabilities in all allowed physical channels are
relevant. It is clear that the bound state must be generated by an interaction
that it is not present in the asymptotic channel, sequestering probability
from a single singlet--singlet color vector from the interaction between
color octets. Such systems are clear examples
of compact four--quark states, in other words, they cannot be expressed in terms of a single physical
channel. Moreover, as can be seen in Table~\ref{t1},
their typical sizes point to compact objects 20\% smaller than a
standard two--meson system.

We have studied the dependence of the probability of a physical channel
on the binding energy. For this purpose we have considered the simplest
system from the numerical point of view, the
$(S_T,I)=(0,1)$ $cc\bar n\bar n$ state. Unfortunately, this state
is unbound for any reasonable set of parameters. Therefore, we bind it by multiplying the
interaction between the light quarks by a fudge factor.
Such a modification does not affect the two--meson threshold while
it decreases the mass of the four--quark state. The results are illustrated in
Fig.~\ref{f2}, showing how in the $\Delta_E\to0$ limit,
the four--quark wave function is almost a pure single physical
channel. Close to this limit one would find what could be defined as
molecular states.
Moreover, the size of the four--quark state increases
when $\Delta_E\to0$.
When the probability concentrates
into a single physical channel ($P_{MM}\to 1$) the
size of the system gets larger than the sum of two isolated mesons~\cite{Vij09}.
We have identified the subsystems responsible for increasing the size of the four--quark state.
Quark-quark ($\langle x^2\rangle^{1/2}$) and antiquark-antiquark ($\langle y^2\rangle^{1/2}$)
distances grow rapidly while the quark--antiquark  distance ($\langle z^2\rangle^{1/2}$)
remains almost constant. This reinforces our previous result, pointing to the appearance
of two-meson-like structures whenever the binding energy goes to zero.
\begin{center}
\includegraphics[width=6cm]{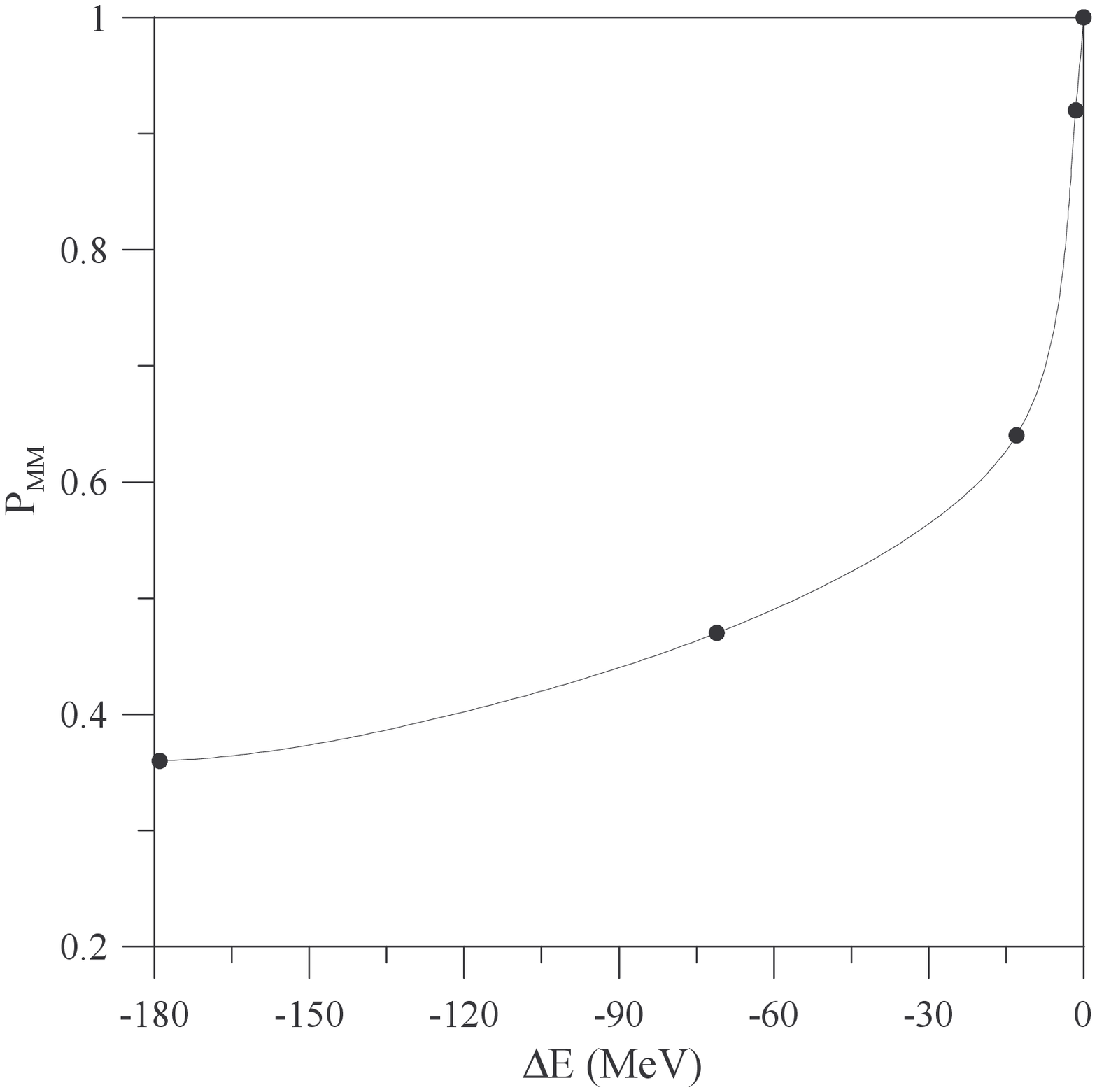}
\figcaption{\label{f2}$P_{MM}$ as a function of $\Delta_E$.}
\end{center}

Although the present analysis has been performed by means of a particular
quark interacting potential~\cite{Vij05}, the conclusions derived are independent 
of the quark-quark interaction
used. They mainly rely on using the same hamiltonian to describe tensors of
different order, two and four-quark components in the present case. When dealing with a
complete basis, any four-quark deeply bound state has to be compact. Only
slightly bound systems could be considered as molecular. Unbound states correspond
to a two-meson system. A similar situation would
be found in the two baryon system, the deuteron could be considered as a 
molecular-like state with a small percentage of its wave function on the $\Delta \Delta$ channel,
whereas the $H-$dibaryon would be a compact six--quark state.
When working with central forces, the only way of getting a bound system is to have
a strong interaction between the constituents that are far apart in the asymptotic limit
(quarks or antiquarks in the present case). In this case the short-range
interaction will capture part of the probability of a two-meson threshold to form a bound
state. This can be reinterpreted as an infinite sum over physical states.
This is why
the analysis performed here is so important before any conclusion can be made concerning 
the existence of compact four--quark states beyond simple molecular structures.

If the prescription of using the same hamiltonian to describe all tensors in the Fock space is relaxed,
new scenarios may appear. Among them, the inclusion of many--body forces is particularly relevant.
In Ref.~\cite{Vij07b} the stability of $QQ\bar n\bar n$ and $Q\bar Q n \bar n$ systems
was analyzed in a simple string model considering only a multiquark confining interaction given
by the minimum of a flip-flop or a butterfly potential in an attempt to discern whether
confining interactions not factorizable as two--body potentials would influence the stability
of four--quark states. The ground state of systems made of two quarks and two antiquarks of
equal masses was found to be below the dissociation threshold. While for the cryptoexotic
$Q\bar Q n\bar n$ the binding decreases when increasing the mass ratio $m_Q/m_n$, for the
flavor exotic $QQ\bar n\bar n$ the effect of mass symmetry breaking is opposite. Others scenarios may emerge
if different many--body forces, like many--body color interactions~\cite{Dmi01} or 't Hooft
instanton--based three-body interactions~\cite{Hoo76}, are considered.

\section{Conclusions}
We have discussed the formalism to express the wave function of a general four--quark state
in terms of physical channels, i.e., those constructed by using color singlet--singlet states.
We have studied charmonium ($c\bar c n\bar n$), bottomonium ($b\bar b n\bar n$)
and exotic ($cc\bar n\bar n$ and $bb\bar n\bar n$)
four-quark states by means of a standard non-relativistic quark potential model.
We look for possible bound states. Among them we are able to distinguish between
meson-meson molecules and compact four-quark states.
The importance of performing a complete analysis of the system, 
energy and wave function, in the vicinity of a two-meson threshold 
has been emphasized.

\end{multicols}

\vspace{-2mm}
\centerline{\rule{80mm}{0.1pt}}
\vspace{2mm}

\begin{multicols}{2}

\end{multicols}

\clearpage

\end{document}